\documentclass[twocolumn,showpacs,preprintnumbers,amsmath,amssymb]{revtex4}

\usepackage{mathbbol}
\usepackage{txfonts}
\usepackage{multirow}
\usepackage{threeparttable}
\usepackage{graphicx}
\usepackage{dcolumn}
\usepackage{bm}

\begin{document}

\preprint{APS/123-QED}

\title{Improved Direct Counterfactual Quantum Communication}


\author{Sheng Zhang}
\email[]{shengzhcn@gmail.com}
\affiliation{Department of Electronic Technology, China Maritime Police Academy, Ningbo 315801, China}
\author{Bo Zhang}
\affiliation{School of Electronic Science and Engineering, National University of Defense Technology, Changsha 410073, China}
\author{Xing-tong Liu}
\affiliation{School of Electronic Science and Engineering, National University of Defense Technology, Changsha 410073, China}


\date{\today}

\begin{abstract}
Recently, a novel direct counterfactual quantum communication protocol was proposed using chained quantum Zeno effect. We found that this protocol is far from being widely used in practical channels, due to the side effect of "chained", which leads to a dramatic increase of the equivalent optical distance between Alice and Bob. Therefore, not only the transmission time of a single bit increases in multiple times, but also the protocol is more sensitive to the noise. Here, we proposed an improved protocol, in which quantum interference is employed to destroy the nested structure induced by "chained" effect. Moreover, we proved that a better counterfactuality is easier to be achieved, and showed that our protocol outperforms the former in the presence of noises.
\end{abstract}

\pacs{03.67.Dd, 03.67.Hk}
\maketitle


\section{Introduction}
\label{Intro}
Quantum communication is now widely accepted to be one of the most promising candidates in future quantum technology. Using quantum mechanics, several amazing tasks, such as dense coding\cite{bennett1992communication,mattle1996dense}, teleportation\cite{bennett1993teleporting,boschi1998experimental} and counterfactual quantum key distribution\cite{Noh09,Sun10}, are naturally achieved. Since the invention of quantum key distribution(QKD) protocol, i.e., BB84 protocol\cite{BB84},  quantum communication has enjoyed a great success with both theoretical and commercial aspects. One of the most significant contributions, which is impossible to be achieved by classical means, is counterfactual quantum communication. It enables two remote parties, Alice and Bob, to exchange messages without transmitting any information carriers.

The idea of counterfactual quantum communication was initialized by interaction-free measurement\cite{EV93,Kwiat99,Noh98}, impressive with the phenomenon that an object can be detected without being intuitively measured. The first example, presented by Noh\cite{Noh09}, was realized in a QKD protocol. Later, we announced a variant adapted to deterministic key distribution scenario\cite{sheng2013counterfactual}. In sharp contrast to conventional QKD schemes, these protocols are counterintuitive that the quantum states, served as the information carriers, never travel through the channel. A translated no-cloning theorem prevents the eavesdroppers from getting any information of the private key. A strict security proof of Noh's protocol(Noh09 protocol) was presented by Yin et al.\cite{Yin10}. We further proved that, although this protocol is secure under a general intercept-resend attack in an ideal mode, the practical security could be compromised due to the dark count rate and low efficiency of the detectors\cite{sheng2012security}. Surprisingly, we also found that Eve could get full information of the key from a real implementation by launching a counterintuitive trojan horse attack\cite{zhang2012counterfactual}. Since the rate of information photons in Noh09 protocol, only up to 12.5\% in ideal setting, is not satisfactory, Sun and Wen improved it to reach 50\% using an iterative module\cite{Sun10}. Experimental verifications of Noh09 protocol have been made by various authors\cite{liu2012experimental,brida2012experimental}.

Most interestingly, the topic of counterfactual quantum communication, has been repainted by Salih et al., who claimed a new protocol(SLAZ2013 protocol) with a better rate, using quantum Zeno effect\cite{salih2013protocol}. They also announced a tripartite counterfactual quantum key distribution protocol\cite{salih2014protocol}, to improve the counterfactuality and security of a preview scheme by Akshata Shenoy H. et al.\cite{shenoy2014counterfactual}. Other interesting applications, such as semi-counterfactual quantum cryptography\cite{shenoy2013semi}, counterfactual quantum-information transfer\cite{guo2014counterfactual}, are also found in recent papers.

Here, we argue that it is problematic to apply SLAZ2013 protocol in real channels, unless the side effect of chained quantum Zeno effect is degraded to an acceptable level. Notice that the equivalent optical distance between Alice and Bob, being amplified by $M\ast N$(numbers of the outer and inner cycles) times, is far larger than the original one, though it is good to use chained quantum Zeno effect to achieve perfect counterfactuality. Consequently, the efficiency, on the first hand, turns out to be very low. In other words, the time taken by transferring a single bit might be much longer than one expects, even though Alice and Bob stand close to each other. On the second hand, the protocol turns out to be more sensitive to the noisy, due to the increase of the equivalent distance. In \cite{salih2013protocol}, it was estimated that an acceptable noise rate only reaches 0.2\%.

In this paper, we present a new quantum counterfactual communication protocol. The rest of the paper is organized as follow: In section\ref{prot}, our protocol is introduced. Then, we analyze the counterfactuality of both our protocol and SLAZ2013 in the following section, and it is showed that our protocol outperforms SLAZ2013 with respect to the counterfactuality and the tolerance of noise. In section\ref{dis}, we have a brief discussion on how to bridge the presented protocol and quantum key distribution. At last, a conclusion is drawn.

\section{Protocol}\label{prot}

First, we give a brief introduction of SLAZ2013 protocol. To achieve the goal of counterfactuality, chained quantum Zeno effect, acting as the core principle, is introduced by employing a series of beam splitters and mirrors. Correspondingly, the optical circuit is divided into two types of cycles, i.e., the outer cycle and inner cycle shown in \cite{salih2013protocol}. At very beginning, a photon, which has nothing to do with the information bit, is injected by the source, and entering the input port of the outer cycle. The rest thing Alice has to do is to observe which of her detectors, $D_{1}$ and $D_{2}$, clicks. At Bob's end, he just chooses to block(pass) the photon, if logic "1"("0") is selected to be transmitted. Let's see how Alice knows the transmitted bit. When "0" is selected, two events, denoted by $E_{1}$ and $E_{2}$ can be observed by Alice:
\begin{itemize}
  \item ($E_{1}$) The photon has been caught in detector $D_{1}$.
  \item ($E_{2}$) The photon has been caught in detector $D_{3}$.
\end{itemize}
Note that $E_{2}$ implies that the photon has been traveling through the channel. Therefore, $E_{2}$ should be discarded. Similarly, when "1" is selected, events $E_{3}$ and $E_{4}$ can be observed:
\begin{itemize}
  \item ($E_{3}$) The photon has been caught in detector $D_{2}$.
  \item ($E_{4}$) The photon has been caught in detector $D_{4}$.
\end{itemize}
Again, $E_{4}$, which goes against the counterfactuality, is discarded.h e central problem is that the twisted structure(i.e., outer cycle twisting with inner cycles) directly increases the optical length of the channel.

\begin{figure}
  \includegraphics{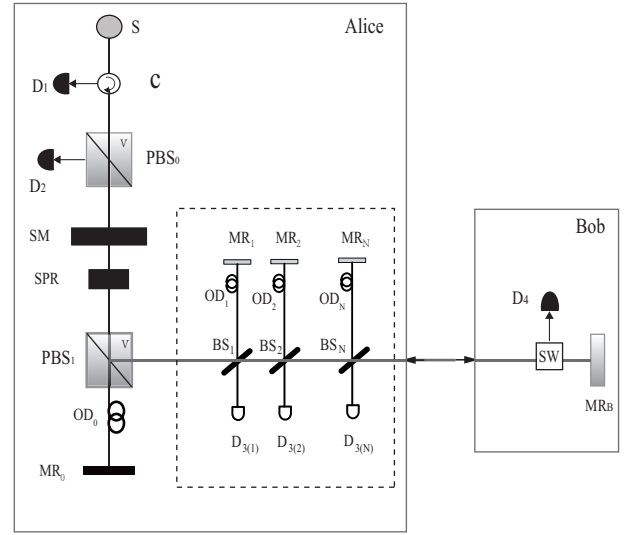}\\
  \caption{Experimental setup. In contrast with SLAZ2013 protocol, an iterative module in the dashed box is introduced to replace the original inner cycle. Here, $BS_{i}$ stands for a beam splitter, and $D_{3(i)}$ denotes a photon detector for $i=$1,2,...,N. Bob uses a switch(SW) to carry out the blocking operation.}\label{setup}
\end{figure}

Now, we introduce our protocol. First, Let's see the setup shown in Fig.\ref{setup}. Compared with SLAZ2013, the only difference is easily found. We put an iterative module(shown in the dashed rectangle), which is as the same as the one in Ref.\cite{Sun10} or \cite{zhang2012counterfactual} except for the mirrors, to replace the inner cycle. Note that the length of each optical delay(OD) in this module should be carefully chosen to match each other. Specifically, the following condition should be satisfied,
\begin{equation}\label{eq1}
    L_{OD_{i+1}}-L_{OD_{i}}=L_{0},
\end{equation}
for $i=1,2,...,N-1$. Here, $L_{OD_{i}}$ and $L_{0}$ denote the optical lengths of $OD_{i}$, and the interval between two neighbouring ODs. Also, $L_{OD_{1}}$ is initialized by the optical length of the real channel in terms of matching.

Next, Let's see how this protocol works. Not surprisingly, the protocol begins with a vertically polarized photon, i.e., the state $|V>$, which will be caught in detector $D_{1}$ or $D_{2}$, according to Bob's choice $1$ or $0$, respectively. The explanation is presented as follow.

\begin{figure}
  \includegraphics{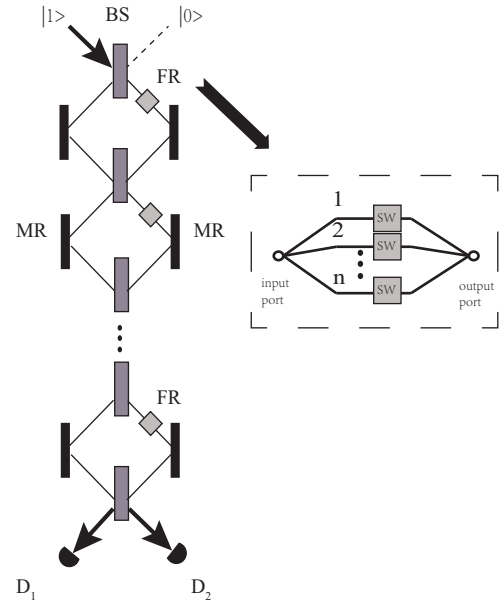}\\
  \caption{Principal schematics. Here FR stands for a fictitious router, explicitly showed in the right side dashed box. The input node routes the photon to the output through one of the $n$ paths.}\label{principle}
\end{figure}

In Fig.\ref{principle}, When Bob passes the photon(all SWs are on), path $i$($i=1,2,...,n$) corresponds to the optical path $PBS_{1}\rightarrow MR_{j}$($j=1,2,...,N$) or $PBS_{1}\rightarrow MR_{B}$. In this case, our protocol degrades to the first step of SLAZ2013, owing to the interference. Therefore, detector $D_{2}$ clicks with certainty. However, when all SWs are off, the interference is destroyed. Consequently, the photon will be caught in the corresponding detector, i.e., $D_{4}$ or $D_{3(i)}$, if it is in the right arm of the BS. In this case, our protocol is also equivalent to the first step of SLAZ2013, excepted that in which detector the photon arrives. In other words, our protocol shares the same principle with SLAZ2013 in spite of some details.

\section{Performance}\label{ana}
In this section, we will focus our attentions on computing the counterfactuality of the presented protocol, and show that the it can be achieved with less resources, in contrast with SLAZ2013 protocol. Moreover, using numerical estimating, we find that our protocol outperforms the former in the presence of channels noises.

Notice that the presented scheme is more efficient than SLAZ2013 protocol with respect to the followings: the equivalent optical distance between Alice and Bob, $D_{eq}$, is only $M*L$, where $L$ denotes the practical distance. Unfortunately, we have $D_{eq}=M*N*L$ for SLAZ2013. Fortunately, the transmission time, i.e., $t=D_{eq}/C$, has been reduced by a factor of $N$.

\subsection{Counterfactuality Analysis}

Next, Let's see how our protocol benefits from the iterative module with respect to counterfactuality. Before the analysis begins, the conception "counterfactuality rate", denoted by $C$,  should be reviewed. Here, it is defined by the probability of a successful communication featured with no transmission of signal carriers. Correspondingly, another conception "abnormality rate" denoted by $A$ is defined by the reverse, so we have $A=1-C$. Now, we are able to define the counterfactuality of a given protocol by a pair of counterfactuality rates(or abnormality rates), $\overrightarrow{C}=$($C_{0}$, $C_{1}$), each representing the counterfactuality rate for Alice sending signal 0 and 1, respectively. Evidently, $C_{i}$($i=0,1$) varies from 0 to 1, and perfect counterfactual communication is available if and only if $C_{i}=1$($i=0,1$).

Back to SLAZ2013 protocol, it is easy to find that her counterfactuality, denoted by $\overrightarrow{C_{1}}$, is given by $P_{1}$ and $P_{2}$, so we have

\begin{equation}\label{eq2}
   \overrightarrow{C_{1}}=(P_{1}, P_{2}).
\end{equation}
Here, $P_{1}$ and $P_{2}$ are given by $P_{1}=|x_{M}|^2$ and $P_{2}=|y_{M,0}|^2$, respectively(Refer to \cite{salih2013protocol} for details). This protocol achieves perfect counterfactuality when $N$ and $M$ approach infinity, leading to $\overrightarrow{C_{1}}\rightarrow(1, 1)$, i.e., $P_{1}=1$ and $P_{2}=1$. Also note that this protocol is realized in two steps, with the following concerning: In the first step, the prototype, referring to Fig.2(a) of Ref.\cite{salih2013protocol}, is only partially counterfactual. In this case, the counterfactuality rate $C_{0}$ turns to be 0 for whatever $N$s. The second step, in which an inner cycle is employed, is introduced to improve the prototype. Therefore, the final protocol makes itself completely sound for counterfactuality. However, in our scheme, we have achieved the same goal by replacing the inner cycle with an iterative module.

Now, we begin to calculate the counterfactuality of our protocol, denoted by $\overrightarrow{C_{2}}$, where $\overrightarrow{C_{2}}=(C_{0}, C_{1})$.  When Bob blocks the channel, it is showed in Fig.\ref{principle} that our protocol is an equivalent transformation of the first step of SLAZ2013.  Therefore, the corresponding rate $C_{1}$ is directly deduced by quantum Zeno effect, i.e.,
\begin{equation}\label{eq4}
   C_{1}=\cos^{2M}\theta,
\end{equation}
equaling to the probability that $D_{1}$ clicks, denoted by $Prob\{D_{1}\ clicks\}$. Similarly, we have $\theta=\pi/2M$. Obviously, perfect counterfactuality is achievable for signal "1", when $M$ approaches infinity.

When Bob passes the photon, i.e., signal 0 is selected, a successful counterfactual communication is established only when there is no photon in the channel for each cycle. Back to Fig.\ref{principle}, the probability that the photon travels through the channel in the $m_{th}$ cycle is given by $t=\prod_{j=1}^{N}t_{j}$, where $t_{j}$ is the transmissivity of the $j_{th}$ BS inside the module. Therefore, $C_{0}$ can be written as
\begin{equation}\label{eq5}
   C_{0}=\prod_{m=1}^{M}(1-\sin^{2}m\theta \cdot t).
\end{equation}

Obviously, Eqs.(\ref{eq4}) and (\ref{eq5}) imply that perfect direct counterfactual communication is seen when $M$ approaches infinity. Interestingly, one obtains $C_{0}=0 $ given that $t=1$, implying that our protocol evolves to the first step of SLAZ2013 protocol as we expected.

\begin{figure}
  \includegraphics{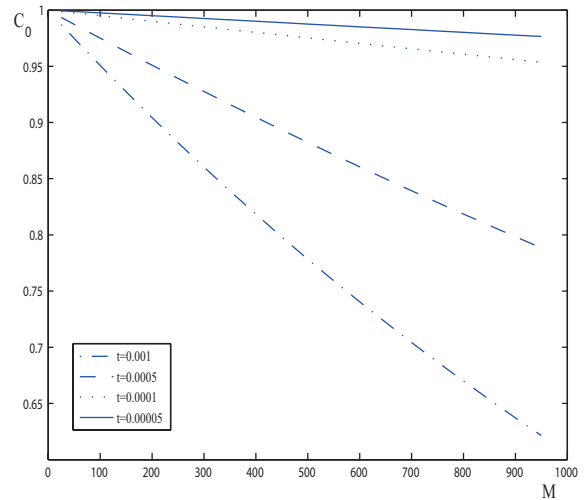}\\
  \caption{$C_{0}$ as a function of $M$.}\label{c0}
\end{figure}

From above analysis, it is known that parameters $N$ and $M$ are crucial to achieve a better performance. We have loosely plotted the counterfactuality rate $C_{0}$, illustrating how it varies as a function of $M$. In Fig.\ref{c0}, it is showed that all curves descend as $M$ increases. In other words, the performance of our protocol becomes worse with a bigger $M$, since the probability that photon exposes itself in the channel evidently increases when the number of the cycles grows up. Fortunately, $C_{0}$ can be improved by reducing $t$(or independently increasing $N$). As is shown in Fig.\ref{c0}, a curve, marked with a smaller $t$, locates itself over the others with bigger ones.  This shows that high counterfactuality (e.g., $C_{0}>0.9$) is achievable with acceptable $M$s, as long as $t$ is chosen to be sufficiently small. In order to show our advantages over SLAZ2013 protocol, we list some meaningful results, obtained from numerical estimating, in table 1. It is clear that, for SLAZ2013 protocol, $N$ should be sufficiently large(meanwhile things getting worse as $M$ increases), in order to achieve acceptable counterfactuality rates(bigger than 0.9). However, the same goal can be achieved, for our protocol, only by choosing an appropriate $t$, keeping $M$ unchanged.

\begin{table}
 \centering
 \caption{Numerical estimating results}\label{tab1}
 \begin{threeparttable}
\begin{tabular}{|c|c|c|c|c|c|c|}
  \hline
  & &$M=25$&$M=50$&$M=75$&$M=100$&$M=150$ \\ \cline{1-7}
  \multirow{4}{*}{(I)\tnote{1}}&$t=0.001$&0.987 &0.975 &0.963 &0.951 &0.927\\ \cline{2-7}
    &$t=0.0005$&0.994& 0.987	& 0.981& 0.975 & 0.963\\ \cline{2-7}
    &$t=0.0001$&0.999	&0.997 &0.996 &0.995 &0.992 \\ \cline{2-7}
    &$t=0.00005$&0.999 & 0.999&0.998&0.997&0.996\\ \cline{2-7}
  \hline
   \multirow{4}{*}{(II)\tnote{2}}&$N=320$&0.912&0.831&0.758&0.693&0.582\\ \cline{2-7}
    &$N=500$&0.943&0.887&0.836&0.788&0.702\\ \cline{2-7}
    &$N=1250$&0.977&0.953&0.930&0.908&0.865 \\ \cline{2-7}
    &$N=2500$&0.988&0.976&0.964&0.953&0.930\\ \cline{2-7}
  \hline
\end{tabular}
 \begin{tablenotes}
  \footnotesize
  \item[1] (I): The first half of the table, corresponding to our protocol. The content units are filled with values of $C_{0}$, referring to different $t$s and $M$s.
  \item[2] (II): The second half of the table, corresponding to SLAZ2013 protocol. The content units are filled with values of $p_{2}$, referring to different $M$s and $N$s.
  \end{tablenotes}
   \end{threeparttable}
\end{table}

Since our protocol shares the same template with the simplified SLAZ2013 protocol, i.e, the first-step protocol, it is easy to conclude the detector rates, $Prob\{D_{1}\ clicks\}$ and $Prob\{D_{2}\ clicks\}$, which are given by $Prob\{D_{1}\ clicks\}=1$ and $Prob\{D_{2}\ clicks\}=\cos^{2M}\theta$.


\subsection{Robustness against Channel Noises}
Here, the robustness of the presented protocol is only investigated in a most representative scenario that the channel noise acts as an obstacle which definitely registers an event of "Block". Errors only occur in case of Bob choosing to pass the photon, where interference is destroyed by noises. Remarkably, for our protocol, the presence of noises definitely induces errors as well as an increase of the probability that detector $D_{3(j)},\ (j=1,2,...,N)$ clicks, which independently discounts the performance of the protocol.

When Bob passes the photon, it will produce a click of detector $D_{2}$ with certainty owing to quantum interference, if the channel is noiseless. Let's see what happens when a "block" in one cycle is triggered by the noise other than Bob. With out loss of generality, we assume that the channel of the $i_{th}$ cycle is blocked due to the noise. Given that the state of the $i_{th}$ cycle is $|\varphi >_{i}=x_{i}|10>+y_{i}|01>$, the quantum state after the $(i+1)_{th}$ BS is written as
\begin{equation}\label{eq6}
   |\varphi >_{i+1}=(x_{i}cos\theta -c*y_{i}sin\theta)|10>+(x_{i}sin\theta+c*y_{i}cos\theta)|01>,
\end{equation}
where $c$ denotes the rate that the single-photon pulse in the channel of the $i_{th}$ cycle is not absorbed. For instance, in SLAZ2013 protocol, $c=0(1)$, corresponding to Bob's choice "Block(pass)", means that the pulse is fully(never) absorbed by Bob's detector $D_{4}$. Interestingly, $c$ varies from 0 to 1 for the presented protocol, since the iterative module also contributes to the benefit that the photon in the right-hand side arm is not absorbed.

\begin{figure*}
  \includegraphics{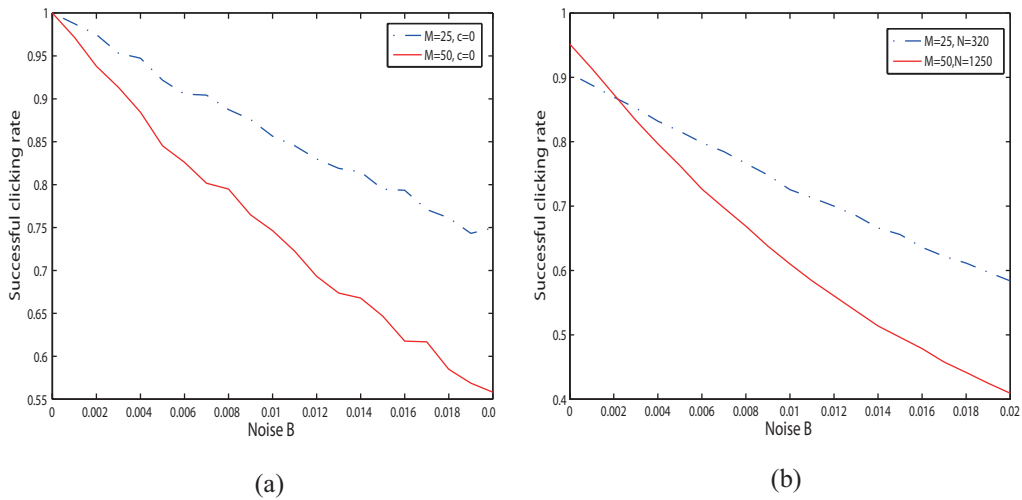}\\
  \caption{Numerical estimating results. (a)Successful clicking rate of our protocol as a function of the noise rate B. Here, $c=0$, corresponding to the worst case, is chosen as a better sample to make a comparison with SLAZ protocol. The solid(dash-dot) curve is plotted for $M=25(50)$. (b)Successful clicking rate of SLAZ2013 protocol as a function of the noise rate B. The solid(dash-dot) curve is plotted for $M=25(50), N=320(1250)$. }\label{noise}
\end{figure*}

Next, it is necessary to fix the rate "c" with given parameters of the module. Suppose that a photon is reflected by $PBS_{1}$ in the $i_{th}$ cycle, it is easy to conclude the probability that it is reflected back to $PBS_{1}$ by one of the mirrors in the module as
\begin{equation}\label{eq7}
    P_{ref}=\sum_{i=0}^{N}\prod_{j=-1}^{i-1}t_{j}^{2}(1-t_{i})^{2},
\end{equation}
due to the absence of quantum interference. Also, the probability that it is absorbed is
\begin{equation}\label{eq8}
    P_{abs}=1-\sum_{i=0}^{N}\prod_{j=-1}^{i-1}t_{j}^{2}(1-t_{i})^{2}-\prod_{i=o}^{N}t_{i}.
\end{equation}
Obviously, we have $c=P_{ref}$. Moreover, it is seen that a photon will be detected by $D_{3(j)},\ (j=1,2,...,N)$ with unit probability, when N approaches infinity\cite{zhang2012counterfactual}. Nevertheless, it is still interesting to investigate the robustness of the presented protocol in finite settings, where $c$ is indeed a non-zero real number less than 1. In order to find how $c$ discounts the rate of detector $D_{2}$, we try to correlate $Prob\{D_{2}\ clicks\}$ with $c$ formally. Assume independently that a single-photon pulse, denoted by an unnormalized quantum state $|\psi >=c|01>$, arrives at the $(i+1)_{th}$ BS in Fig.\ref{principle}, the final state after the following $M-i$ BSs is written by
\begin{equation}\label{eq9}
    |\psi >_{final}=c*(cos(M-i)\theta |01>-sin(M-i)\theta |10>),
\end{equation}
owing to quantum interference. Obviously, it is seen from Eq.\ref{eq9} that this independent pulse definitely contributes to the rate of detector $D_{2}$. Therefore, $Prob\{D_{2}\ clicks\}$, which is given by $Prob\{D_{2}\ clicks\}=(1-(1-c)y_{i}cos(M-i)\theta)^2$, directly increases as $c$ increases. Eq.\ref{eq7}
shows the balance of $c$ and $N$, and that $c=0$ when $N\rightarrow \infty$ for the worse case. Now, we use the same technique, that random numbers between 0 and 1 are employed to play the role of noise, to plot the successful rate of the right detector. The curves in Fig.\ref{noise} (a) are statistically averaged on 2000-times repetition to achieve better smoothness. Obviously, our protocol outperforms SLAZ2013 protocol on the tolerance of noise given the same $M$. Moreover, a smaller $M$ implies a bigger tolerance of $B$ for our protocol, which is consistent with the fact that an increase of the number of cycles immediately increases the risk of suffering from channel noises.

\section{Discussions}\label{dis}
Although both our protocol and SLAZ2013 provide us ways to establish counterfactual communication, which is impossible with classical means, we should point out that they always fail in a secure scenario, such as quantum key distribution. The central problem is that no-cloning theorem is not included in their principles. Specifically, in both protocols, only orthogonal states, say, $|\phi_{0}>$ and $|\phi_{1}>$, are employed. Fortunately, it is not difficult to make them secure. All one should do is to change the pure states into nonorthogonal mixed states, i.e., $Tr[\rho_{0} \rho_{1}]\neq0$. For this, Noh09 protocol acts as a good example. In doing so, our protocol immediately evolves to a quantum key distribution scheme. Here, we also highlight an open question that whether it is possible to explore unconditional security directly from quantum Zeno effect, thus leading to a new paradigm outperforming existed quantum key distribution schemes.

\section{Conclusion}\label{con}
It is interesting that direct counterfactual quantum communication is achievable using quantum Zeno effect. However, the original scheme, i.e., SLAZ2013 protocol, has new problems when applying it in real channels. First, the efficiency is low, compared with conventional schemes. Second, it is too sensitive to  noise. We find that those two flaws are resulted from the nested structure, i.e., the inner cycles and outer cycles. In this paper, we succeeded in reducing the cycles by replacing the inner cycle with an iterative module, which is the core component of our new protocol. We have proved that perfect counterfactuality is achievable for our protocol, and showed that the a given level of performance can be reached with less cycles. Next, we further discussed the robustness of our protocol, and numerical estimating results showed that our scheme outperforms SLAZ2013 in the presence of noise. At last, we discussed how to bridge our protocol and quantum key distribution with no-cloning theorem, in order to broaden the view of direct counterfactual quantum communication.

\section{Acknowledgements}
This work was supported by National Natural Science Fundation of China with the project number 61300203.

\bibliography{my_bib_database}

\end{document}